\documentstyle[twocolumn,aps,epsfig,floats]{revtex}
\begin{document}

\twocolumn[\hsize\textwidth\columnwidth\hsize\csname %
@twocolumnfalse\endcsname

\title{Magnetic Coherence as a Universal Feature of Cuprate Superconductors}
\author{Dirk K. Morr $^{1}$ and David Pines $^{2}$}
\address{ $^{1}$ Theoretical Division, Los Alamos National
Laboratory, Los Alamos, NM 87545\\
$^{2}$ Institute for Complex
Adaptive Matter, University of California, and LANSCE-Division,
Los Alamos National Laboratory, Los Alamos, NM 87545}
\date{\today}
\draft
\maketitle
\begin{abstract}
Recent inelastic neutron scattering (INS) experiments on
La$_{2-x}$Sr$_x$CuO$_4$ have established the existence of a {\it magnetic
coherence
effect}, i.e., strong frequency and momentum dependent changes of
the spin susceptibility, $\chi''$, in the superconducting phase.
We show, using the spin-fermion model for
incommensurate antiferromagnetic spin fluctuations, that the
magnetic coherence effect establishes the ability of INS experiments to probe
the
electronic spectrum of the cuprates, in that the
effect arises from the interplay of an incommensurate magnetic
response, the form of the underlying Fermi surface, and the opening
of the d-wave gap in the fermionic spectrum. In particular, we find
that the magnetic coherence effect observed in INS experiments on
La$_{2-x}$Sr$_x$CuO$_4$ requires that the Fermi surface be closed around
$(\pi,\pi)$ up to optimal doping. We present several
predictions for the form of the magnetic coherence effect in
YBa$_2$Cu$_3$O$_{6+x}$ in which an incommensurate magnetic
response has been observed in the superconducting state.
\end{abstract}

]

\narrowtext

\section{Introduction}
The spin excitation spectrum in La$_{2-x}$Sr$_x$CuO$_4$ (LSCO)
\cite{Shi89,Mas96,Aep97,Yam98,Lake99} and YBa$_2$Cu$_3$O$_{6+x}$
(YBCO) \cite{Tra92,Dai96,Dai97,Fong97,Dai98} in the normal and
superconducting (SC) state has been intensively studied by
inelastic neutron scattering (INS) experiments over the last few
years. While it has been known for some time that the magnetic
response in LSCO compounds with $x>0.04$ is characterized by peaks
in $\chi''({\bf q}, \omega)$ at incommensurate wave-vectors ${\bf
Q}_i=(1 \pm \delta,1) \pi$ and ${\bf Q}_i=(1,1 \pm \delta) \pi$
\cite{Shi89,Mas96,Yam98}, an incommensurate structure in several
YBCO compounds has only recently been reported
\cite{Tra92,Dai97,Mook98}. Recent INS experiments on
La$_{2-x}$Sr$_{x}$CuO$_4$ by Mason {\it et al.}~(x=0.14)
\cite{Mas96} and Lake {\it et al.}~(x=0.16) \cite{Lake99} have established the
presence of a magnetic coherence effect --
strong momentum and frequency dependent changes in $\chi''$ when
entering the superconducting state. For both doping
concentrations, $\chi''({\bf Q}_i, \omega)$ in the superconducting state
is considerably decreased from its normal state value below
$\omega \approx 7$ meV, while it increases above this frequency.
For frequencies in the vicinity of 7 meV, the incommensurate peaks
sharpen in the superconducting state, while at higher frequencies
the peak widths in the normal and superconducting state are
approximately equal.

We have recently shown\cite{Morr00a} that the magnetic coherence
effect in LSCO is a direct consequence of changes in the damping of
incommensurate antiferromagnetic spin fluctuations due to the
appearance of a d-wave gap in the fermionic spectrum in the superconducting
state. Our theoretical results, based on the
spin-fermion model for incommensurate antiferromagnetic
spin-fluctuations, for the frequency and
momentum dependent changes of $\chi''$ in the superconducting
state are in good qualitative, and to a large extent
quantitative, agreement with the experimental data.  In the present
communication, we extend our earlier work. We present novel
predictions for the magnetic coherence effect in parts of the
magnetic Brillouin zone (BZ) so far unexplored by INS experiments, and
 provide further support for the arguments presented in
Ref.~\cite{Morr00a} that INS data in the superconducting state
provide information on the symmetry of the order parameter and the
topology of the Fermi surface (FS); in particular, we discuss our
results for $\chi''$ in those parts of the BZ where the so-called
spin-gap vanishes. We then extend our calculations to YBCO, where
incommensurate peaks have quite recently been found in the superconducting
state, and make detailed
predictions for the magnetic coherence effect in YBCO. To the
extent that our predictions are confirmed by future experiments,
our work will demonstrate that INS experiments are {\it not}
confined to providing information on the magnetic excitation
spectrum, but can also probe the {\it
electronic spectrum} in the superconducting state.

Our paper is organized as follows. In
Sec.~\ref{theory} we briefly describe our theoretical model. In
Secs.~\ref{LSCO} and \ref{YBCO} we present our theoretical results
for the magnetic coherence effect in LSCO and YBCO, and, where
possible, compare these with experimental data. In Sec.~\ref{concl}
we summarize our results and discuss their implications for future experiments.

\section{Theoretical Model}
\label{theory}

The starting point for our calculations is a spin-fermion model
\cite{sfmodel} for incommensurate antiferromagnetic spin
fluctuations. In this model spin-excitations interact with
low-energy  fermionic quasi-particles via
\begin{equation}
{\cal H}_{sf}= -g \sum_{q,k} {\bf S}_q c^\dagger_{\alpha, k-q}
{\bf \sigma}_{\alpha,\beta} c^\dagger_{\beta, k} \label{Hsf}
\end{equation}
where $g$ is the unrenormalized the spin-fermion coupling, and
${\bf \sigma}_{\alpha,\beta}$ are the Pauli-matrices. The spin
propagator, $\chi$, is renormalized by the interaction with the
fermionic  degrees of freedom and given by
\begin{equation}
\chi^{-1} = \chi_0^{-1} -  \Pi  \ ,
\label{Dyson}
\end{equation}
where $\chi_0$ is the bare propagator, and $\Pi$ is the bosonic
self-energy given by the irreducible particle-hole bubble. Within
this model $\chi_0$ is obtained by integrating out the high-energy
fermionic degrees of freedom. However, since a reliable
description of fermionic excitations at high frequencies is not
yet possible, a microscopic calculation of $\chi_0$ is not yet
feasible. We therefore make the experimentally motivated ansatz
that $\chi_0$
takes the form
\begin{equation}
\chi_0^{-1}= { \xi_0^{-2} + ({\bf q} - {\bf Q}_i)^2 \over
\alpha } \ ,
\label{chi0}
\end{equation}
where $\xi_0$ is the {\it bare}  magnetic correlation
length and $\alpha$ is a temperature independent constant. In
general one would expect a frequency-dependent term in Eq.(\ref{chi0}); it
is omitted here because experimentally there is no observed
dispersion in the spin excitation spectrum for the frequency range
considered below \cite{Maspc}.
With the ansatz, Eq.(\ref{chi0}), we note that in Eq.(\ref{Dyson})
$\chi_0$ specifies the momentum dependence of the
incommensurate peaks in the absence of coupling to the particle-hole
excitations described by $\Pi$, while any frequency dependence of $\chi''$ is
solely determined by $\Pi$. In what follows it is convenient to
introduce a renormalized magnetic correlation length,
\begin{equation}
\xi^{-2} = \xi_0^{-2} - \alpha \, {\rm Re} \, \Pi \ ,
\end{equation}
to take into account the influence of particle-hole excitations on the static
susceptibility.

The frequency and momentum
structure of the magnetic coherence effect can be obtained
by considering the lowest order diagrams (in $g$) in the
expansion of $\Pi$. In the normal state, one has
\begin{eqnarray}
\Pi_{N}({\bf q}, i \omega_n) &=& -g^2 \, T \sum_{{\bf k},m} \
G({\bf k}, i\nu_m) \\
& & \qquad \times G({\bf k+q}, i\nu_m+i\omega_n)
\label{PiN}
\end{eqnarray}
where $G^{-1}({\bf k}, i\nu_m)=i \nu_m - \epsilon_k$ is the
fermionic Greens function. The normal state
electronic tight-binding dispersion, $\epsilon_k$, is given by
\begin{eqnarray}
\epsilon_{\bf k} &=& -2t \Big( \cos(k_x) + \cos(k_y) \Big)
\nonumber \\ & & \quad -4t^\prime \cos(k_x) \cos(k_y)  -\mu \ ,
\label{dispersion}
\end{eqnarray}
where $t, t^\prime$ are the hopping elements between nearest and
next-nearest neighbors, respectively, and $\mu$ is the chemical
potential.
The structure of $\Pi$ changes in the superconducting state, being
given by
\begin{eqnarray}
\Pi_{SC}({\bf q}, i \omega_n) &=& -g^2 \, T \sum_{{\bf k},m} \
\Big\{ G({\bf k}, i\nu_m) G({\bf k+q}, i\nu_m+i\omega_n) \nonumber
\\ & & + F({\bf k}, i\nu_m) F({\bf k+q}, i\nu_m+i\omega_n)
\Big\} \ ,
\label{PiSC}
\end{eqnarray}
where $G$ and $F$ are the normal and anomalous Green's functions
\begin{eqnarray}
G&=&{ i\omega_n + \epsilon_k \over (i \omega_n)^2 -\epsilon_k^2 -
\Delta_k^2 }, \ \ F={\Delta_k \over (i \omega_n)^2 -\epsilon_k^2 -
\Delta_k^2 } \ ,
\end{eqnarray}
$E_{\bf k}= \sqrt{ \epsilon_{\bf k}^2 + |\Delta_{\bf k}|^2}$ is
the fermionic dispersion in the superconducting state, and
\begin{equation}
\Delta_{\bf k}=\Delta_{0} { \cos(k_x) - \cos(k_y) \over 2 }
\label{dwave}
\end{equation}
is the d-wave gap.

Before presenting our theoretical results in the next section, we
discuss briefly the range of applicability of the model introduced
above. A close comparison of nuclear magnetic resonance (NMR) experiments (a
local probe) and INS experiments (a global probe) leads one to the conclusion
that incommensuration is a global property of the cuprates, brought about by
the formation of dynamic magnetic domain walls \cite{Morr00b}. Within each
magnetic domain, the spin fluctuations are commensurate. As we have shown in Ref.~\cite{Morr00a} and discuss
further in the next
section, the magnetic coherence effect thus far observed
arises predominantly from
``cold" quasi-particle transitions between regions of
the FS close to the
superconducting nodes (excitations 1 and 2 in
Fig.~\ref{exc}).
Hence, for our model to be valid, these cold quasiparticles
must see the incommensuration which implies that their mean free path must be
large compared to the domain size of $O(\xi)$, a condition which is satisfied
experimentally \cite{Sto96}. Moreover, the charge density variation associated
with the intrinsically inhomogeneous behavior must be sufficiently small
that the planar quasi-particle momentum, ${\bf k}$, continues to be a good
quantum number; recent NMR experiments \cite{Haase} suggest that this is the
case.

A further important point concerns the validity of the description of the
fermionic
degrees of freedom in the superconducting state by BCS Greens functions.
ARPES experiments in under-, optimally and overdoped HTSC
compounds have shown that the electronic spectral function of the cold
quasi-particles can be reasonably well approximated by its
BCS form \cite{arpes}. However, our calculations predict an as yet
unobserved feature of the magnetic coherence effect, a sharp drop
in $\chi''$ for frequencies $\omega \approx 2 \Delta_0$, that
originates in ``hot" quasi-particle excitations located in regions around
$(0,\pi)$ and symmetry related points (excitations 3 and 4 in
Fig.~\ref{exc}). In these regions of the FS, particularly in the
underdoped HTSC compounds, the formation of a leading edge gap in the normal
state causes the electronic spectral function
to
deviate greatly from its BCS form \cite{Chu98,arpes1}; such deviation
might prevent the observation of the predicted sharp decrease in
$\chi''$ at higher energies. We will return to this issue when we
discuss the magnetic coherence effect in underdoped YBCO
compounds.

The model introduced above is thus valid as long as the fermionic
quasi-particles in the vicinity of the superconducting nodes {\bf
(a)} possess a sufficiently large mean free path, and {\bf (b)}
can be reasonably well described by a BCS Greens function. Our
model should thus possess maximum applicability for the optimally and overdoped
HTSC compounds, while experiments on the underdoped compounds will disclose
whether significant deviations arise as a result of the leading edge gap.

\section{Theoretical Results and Comparison with Experiments}

\subsection{Magnetic Coherence in La$_{2-x}$Sr$_x$CuO$_4$}
\label{LSCO}

A brief account of
our results, based on the model introduced above, has already been presented in
Ref.~\cite{Morr00a}, where which we showed it provides a
qualitative and to a large
extent quantitative explanation for the magnetic coherence effect in LSCO.
However, we
believe that it is beneficial for the general understanding and
discussion of our approach and its extension to thus far unexplored regions of
the Brillouin zone, to review briefly some of
these earlier results.

We consider LSCO with $x=0.16$, for which the parameter set
$t^\prime/t=-0.22$
and $\mu/t=-0.84$ yields a closed Fermi surface around $(\pi,\pi)$, in agreement
with the
experimental data on both the magnetic coherence effect and the
subsequent ARPES experiments by Ino {\it et al.}~\cite{Ino98} on
a variety of LSCO compounds with different hole concentrations. The
superconducting gap, $\Delta_{0}\approx 10$
meV, was taken from the analysis of
Raman scattering experiments by
Chen {\it et al.}~\cite{Chen94}, while the
incommensurate
wave-vector ${\bf Q}_i$ is at $\delta \approx 0.25$
\cite{Mas96,Yam98}.

\begin{figure} [t]
\begin{center}
\leavevmode
\epsfxsize=7.5cm
\epsffile{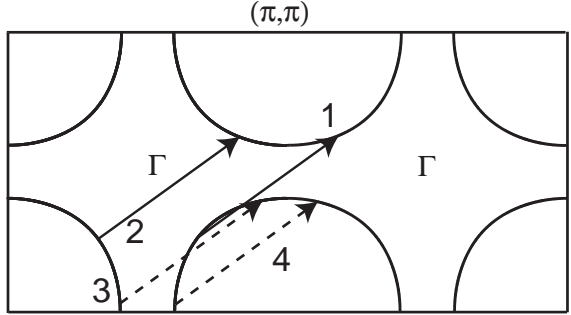}
\end{center}
\caption{Fermi surface of La$_{2-x}$Sr$_{x}$CuO$_4$ and
particle-hole excitations with wave-vector ${\bf
Q}_i$.}
\label{exc}
\end{figure}

\begin{figure} [t]
\begin{center}
\leavevmode
\epsfxsize=7.5cm
\epsffile{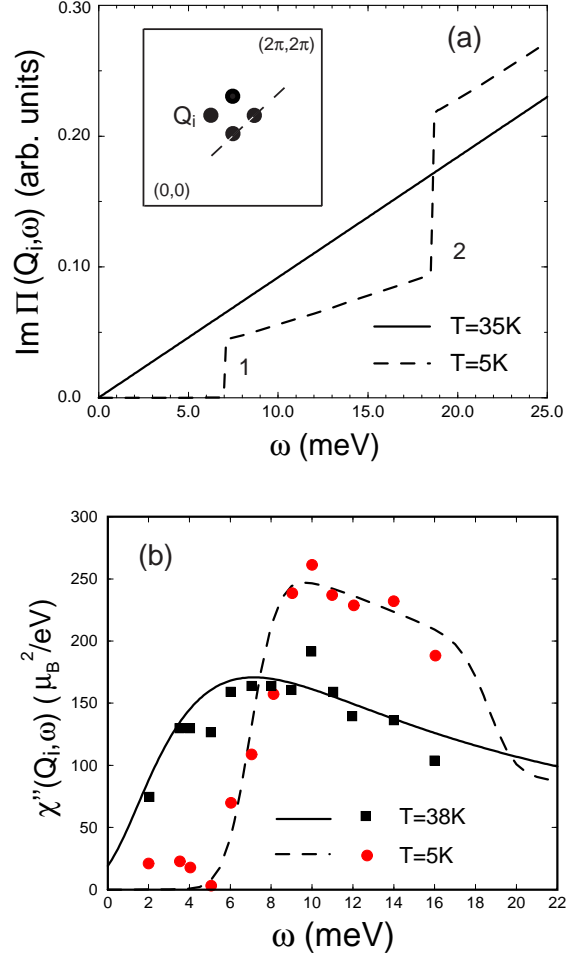}
\end{center}
\caption{ {\it (a)} The spin-damping Im$\, \Pi$ at ${\bf Q}_i$ as
a function of frequency in the normal (solid line) and
superconducting state (dashed line) of
La$_{1.86}$Sr$_{0.14}$CuO$_4$. {\it (b)} Fit of our theoretical
results for $\chi''({\bf Q}_i, \omega)$ (lines) to the
experimental data Ref.\protect\cite{Lake99} (filled circles and
squares) in the normal (solid line) and superconducting state
(dashed line).}
\label{sd_om}
\end{figure}

We consider first the frequency dependence of $\chi''$ at
${\bf q}={\bf Q}_i$ in the normal state.  The particle-hole
excitations that at low frequencies predominantly contribute to
the spin-damping, Im$\, \Pi$, connect two points on the FS
separated by the wave-vector ${\bf Q}_i$. Since ${\bf Q}_i$ is
incommensurate, the four possible decay channels for spin
excitations are those shown in Fig.~\ref{exc}.
In the normal state all four channels are excited for $\omega \not
= 0$ and it follows from Eq.(\ref{PiN}) that Im$\, \Pi_{N} =
\gamma \omega$ \cite{Chu97} where $\gamma$ decreases with
increasing temperature. The linear frequency dependence of Im$\,
\Pi$ that we present in Fig.~\ref{sd_om}a (solid line) is a
general result for all momenta which connect two points on the FS.
Explicit calculations furthermore yield Re$\, \Pi_{N} \approx
const.$~over a large frequency range, and thus
$\xi_{N}^{-2}(\omega) = const$. The resulting dynamic
susceptibility
\begin{equation}
\chi''({\bf Q}_i, \omega) = {  \alpha^2 \gamma \omega \over
\xi^{-2} + (\alpha \gamma \omega)^2 } \ ,
 \label{chiN}
\end{equation}
is of the MMP form \cite{MMP}, and quantitatively describes the results of
INS experiments in the normal state of LSCO and YBCO
\cite{Morr00b}.

We now turn to our results, presented in Fig.~\ref{sd_om}a, for
Im$\, \Pi_{SC}$ at ${\bf Q}_i$ in the SC state, Eq.(\ref{PiSC}).
 Since the fermionic dispersion acquires a
d-wave gap in the SC state, the four channels for quasi-particle
excitations split into two pairs with degenerate non-zero
threshold energies, $\omega_c^{(1,2)}$, that are determined by
\begin{equation}
\omega_c^{(1,2)}=|\Delta_{\bf k}|+|\Delta_{\bf k+Q_i}| \ .
\end{equation}
Here both ${\bf k}$ and ${\bf k+Q}_i$ lie on the Fermi surface, as
shown in Fig.~\ref{exc}. For the band parameters chosen, the
threshold energies are $\omega_c^{(1)}=0.70 \Delta_0$ for
quasi-particle excitations close to the nodes of the
superconducting gap (excitations 1 and 2), and $\omega_c^{(2)}=1.86
\Delta_0$ for excitations that connect momenta around $(0,\pi)$
and $(\pi,0)$ (excitations 3 and 4). Thus, $\omega_c^{(1,2)}$ is clearly
related to the momentum dependence of the order parameter and the
shape of the Fermi surface. Since for $T=0$ and
$\omega<\omega_c^{(1)}$, Im$\, \Pi_{SC} \equiv 0$, and thus
$\chi_{SC}'' \equiv 0$, $\omega_c^{(1)}$ is often referred to as
the spin-gap in the superconducting state. Furthermore, the
superconducting coherence factors in Eq.(\ref{PiSC}) determine how
Im$\, \Pi_{SC}$ increases above $\omega_c^{(1,2)}$. Since excitations
1-4 connect parts of the FS where the superconducting gap,
$\Delta_k$, possesses different signs, Im$\, \Pi_{SC}$ exhibits
jumps at $\omega_c^{(1,2)}$. Were the superconducting gap
to possess $s-$wave symmetry, $\omega_c^{(1)}=\omega_c^{(2)}=2
\Delta_0$, this would imply that $\chi_{SC}'' \not = 0$ only for
frequencies above $2 \Delta_0$. Furthermore, since in this case
the quasi-particle excitations connect regions with the same sign
of $\Delta_k$, Im$\, \Pi_{SC}$ would increase continuously above
$2 \Delta_0$. The experimentally observed sharp increase in
$\chi_{SC}''$ (see the experimental data in Fig.~\ref{sd_om}b)
above $\omega_c^{(1)}$ provides strong support for a sign change
of $\Delta_k$ across the FS, and thus for a d-wave symmetry of the
gap.

While Im$\, \Pi_{SC}$ is dominated by quasi-particle excitations
that are confined to the Fermi surface, particle-hole excitations
in the whole BZ contribute to Re$\, \Pi_{SC}$. Since a description
of the fermions away from the FS by a simple BCS ansatz for the
Greens function is not adequate, a detailed calculation of Re$\,
\Pi_{SC}$ even to lowest order in $g$ is not yet feasible.
Moreover, while a naive calculation using Eq.(\ref{PiSC}) shows
that Re$\, \Pi_{SC}$ exhibits logarithmic divergences at the
threshold frequencies $\omega_c^{(1,2)}$ for $T=0$ due to the
steps in Im$\, \Pi_{SC}$, it has recently been shown \cite{Chu98} that
these are an artifact of our restriction to the second order
bosonic self-energy correction. When fermionic lifetimes are
calculated within a self-consistent strong-coupling approach, the
steps in Im$\, \Pi_{SC}$ are smoothed out for $T \not = 0$, which
implies that the weak logarithmic divergences in Re$\, \Pi_{SC}$
become a smooth function of frequency. Since the spin-gap below a
frequency $\approx \omega_c^{(1)}$ survives the inclusion of
realistic fermionic lifetimes \cite{Chu98} we expect
that the conclusions we draw in the following are valid beyond the
current level of approximation.

In Fig.~\ref{sd_om}b we present a fit of our theoretical results
for $\chi''$ to the experimental data of Ref.~\cite{Lake99} in the
normal and superconducting state. The fit to the experimental data
in  the superconducting state for $\omega_c^{(1)}<\omega<16$ meV
was obtained by making the ansatz that $\xi_{SC}(\omega)$ is
frequency independent in this range and given by
\begin{equation}
\xi^{2}_{SC}(\omega) \approx  { 3 \over 2} \, \xi^{2}_N = const.
\label{xi_sc}
\end{equation}
We take the experimental energy resolution into account by
convoluting our theoretical results with a Gaussian distribution
of width $\sigma \approx 2$ meV. As may be seen in
Fig.~\ref{sd_om}b, the simple ansatz, Eq.(\ref{xi_sc}), yields
good agreement with experiment.

\begin{figure} [t]
\begin{center}
\leavevmode
\epsfxsize=7.5cm
\epsffile{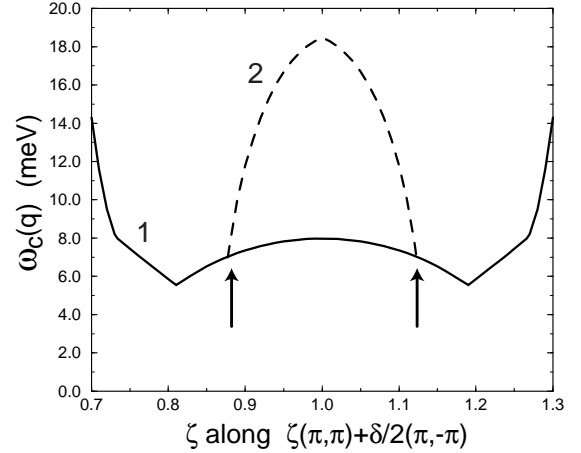}
\end{center}
\caption{ Momentum dependence of the two lower frequency thresholds arising from
excitations 1 (solid line) and 2 (dashed line) along
the path shown in  the inset of Fig.~\protect\ref{sd_om}a. The arrows indicate
the position of
the incommensurate peaks at $\zeta=1\pm \delta/2=0.88 (1.12)$.}
\label{om_c1_c2}
\end{figure}

The approximate frequency independence of $\xi_{SC}$,
Eq.(\ref{xi_sc}), is a consequence of the redistribution in the
spectral weight of $\chi''({\bf Q}_i, \omega)$ in the
superconducting state. In general it should be possible to arrive
at a self-consistent description of the frequency dependence of
$\xi_{SC}$ by applying the Kramers-Kronig relation to
$\chi_{SC}''(\omega)$. We find, however, that $\xi_{SC}(\omega)$
for $\omega<\omega_c^{(2)}$ is sensitive to the form of
$\chi_{SC}''(\omega)$ above $\omega_c^{(2)}$. Since this form is
at moment not well known, a self-consistent description of
$\xi_{SC}$ is beyond the scope of this paper.

\begin{figure} [t]
\begin{center}
\leavevmode
\epsfxsize=7.5cm
\epsffile{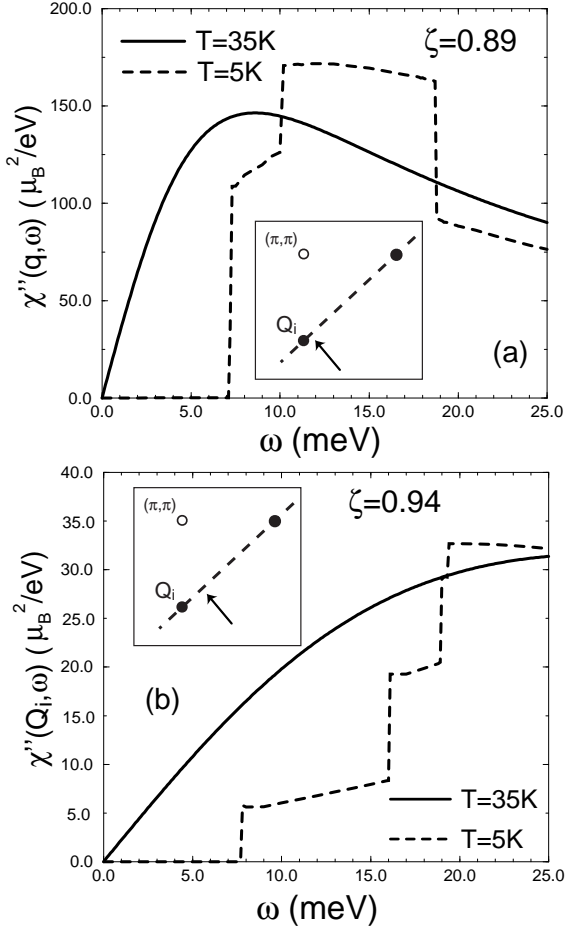}
\end{center}
\caption{ $\chi''({\bf q}, \omega)$ in the normal (solid line) and
superconducting  state (dashed line) for {\it (a)} $\zeta=0.89$
and {\it (b)} $\zeta=0.94$.} \label{zeta}
\end{figure}
Another interesting feature of the magnetic coherence effect is
the change in the frequency dependence of $\chi_{SC}''$ as one
moves away from the incommensurate peak position. To demonstrate
this we consider the frequency dependence of $\chi_{SC}''$ for
several momenta along the path ${\bf q}=\zeta
(\pi,\pi)+\delta/2(\pi,-\pi)$ that connects two incommensurate
peaks and is shown in the inset of Fig.~\ref{sd_om}a. We first
note that due to the symmetry of the FS, the double-degeneracy of
the threshold frequencies is lifted for ${\bf q} \not = {\bf
Q}_i$, resulting in four different threshold energies for the
quasi-particle excitations shown in Fig.~\ref{exc}. In
Fig.~\ref{om_c1_c2} we show the momentum dependence of the two
lower threshold frequencies along the momentum path shown in the
inset in Fig.~\ref{sd_om}a. The lowest threshold frequency (solid
line), which corresponds to the spin-gap, originates from
excitation (1) in Fig.~\ref{exc}, whereas the next higher
threshold (dashed line) stems from excitation (2). The threshold
frequency of excitation (2) increases rapidly as one moves away
from ${\bf Q}_i$ towards the center of the scan.  In the opposite
direction, excitation (2) terminates at a momentum very close to
${\bf Q}_i$ for which it becomes impossible to connect two points
on the FS.

The splitting of the threshold frequencies is clearly observable
in Fig.~\ref{zeta}, where we present the frequency dependence of
$\chi_{SC}''$ for two different values of $\zeta$. The arrow in
the insets shows the respective positions in momentum space. While
the frequency separation of the two upper thresholds is small for
$\zeta=0.89$ (Fig.~\ref{zeta}a), the energy gap between the two
lower threshold energies increases rapidly with increasing
distance from ${\bf Q}_i$; this leads to two sharp increases of
$\chi_{SC}''$ at low frequencies. The frequency dependence of
$\chi_{SC}''$ changes even further with larger distance from ${\bf
Q}_i$, as is shown in Fig.~\ref{zeta}b for $\zeta=0.94$, where
$\chi_{SC}''$ now sharply increases at all four (!) threshold
energies. This result follows straightforwardly from
Eqs.(\ref{Dyson}) and (\ref{chi0}) since $\chi_{SC}''({\bf q},
\omega)$ exhibits a maximum at a frequency where
\begin{equation}
\xi^{-2}+({\bf q-Q}_i)^2 = \alpha {\rm Im} \Pi({\bf q}, \omega)
 \ .
 \label{max}
\end{equation}
Since the overall scale of Im$\, \Pi_{SC}$ does not increases as
one moves away from ${\bf Q}_i$, it follows that the maximum of
$\chi_{SC}''$ shifts to higher frequencies, in agreement with the
results in Fig.~\ref{zeta}b. We thus predict that as one moves
away from ${\bf Q}_i$, $\chi_{SC}''$ acquires a new frequency
structure originating from the lifting of the degeneracy of the thresholds
energies. In particular, the decrease of $\chi_{SC}''$ at
$\omega_c^{(2)}$ transforms into an increase in the intensity away
from the incommensurate position.

\begin{figure} [t]
\begin{center}
\leavevmode
\epsfxsize=7.5cm
\epsffile{Fig5.ai}
\end{center}
\caption{Momentum dependence of the spin-gap, i.e., the lowest
frequency threshold, along the path shown in the inset of
Fig.~\protect\ref{sd_om}a for two different values of $t^\prime$.
The arrows indicate the positions of the incommensurate peaks.}
\label{om_c1_comp}
\end{figure}
Since the spin-gap is determined by the minimum energy for
excitation (1) (Fig.~\ref{exc}) that connects two points on the
FS close to the nodes of the superconducting gap, it turns out to
be sensitive to changes in the form of the FS.  To demonstrate
this sensitivity we plot the spin-gap for two different values of
$t^\prime$ in Fig.~\ref{om_c1_comp}. We find that as
$|t^\prime/t|$ is increased, the spin-gap shifts up in energy,
while the location of its minimum at $q_{min}$ shifts away from
the center of the scan. Since the momentum dependence of the
spin-gap varies strongly even for small changes in the FS, its
experimental observation provides valuable constraints for
the form of the underlying FS.

A special momentum, ${\bf q}_n$, in the magnetic BZ is that which
connects the nodes of the superconducting gap, since here the
spin-gap vanishes and one expects to find gapless spin
excitations. It follows from the form of the FS shown in
Fig.~\ref{exc}, that there are three different momenta for gapless
excitations. The momentum, ${\bf q}_n$, for which we expect the
largest intensity in $\chi''$, both in the normal and superconducting state,
lies along the diagonal of the magnetic BZ.
\begin{figure} [t]
\begin{center}
\leavevmode
\epsfxsize=7.5cm
\epsffile{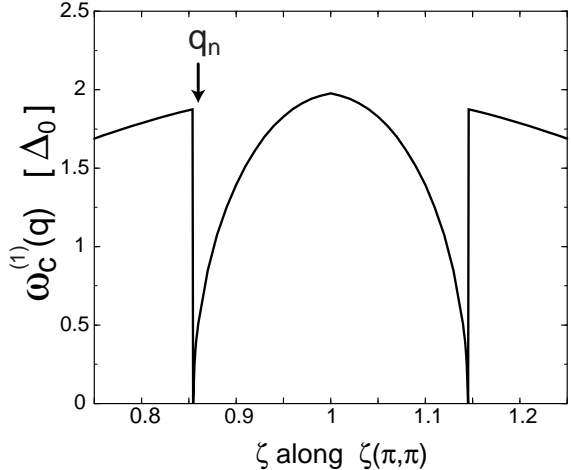}
\end{center}
\caption{Momentum dependence of the spin-gap along the diagonal of
the magnetic BZ (see inset of Fig.~\protect\ref{chi_qp}).}
\label{sg_q}
\end{figure}
We therefore present in Fig.~\ref{sg_q} the momentum dependence of
the spin-gap along the path ${\bf q} = \zeta (\pi,\pi)$ and find
as expected that the spin-gap vanishes at $\zeta=0.855$
corresponding to a wavevector that connects the nodes of the
superconducting gap. Note that the spin-gap rises steeply in the
vicinity of ${\bf q}_n$ due to the large Fermi velocity along the
diagonal of the fermionic BZ.

\begin{figure} [t]
\begin{center}
\leavevmode
\epsfxsize=7.5cm
\epsffile{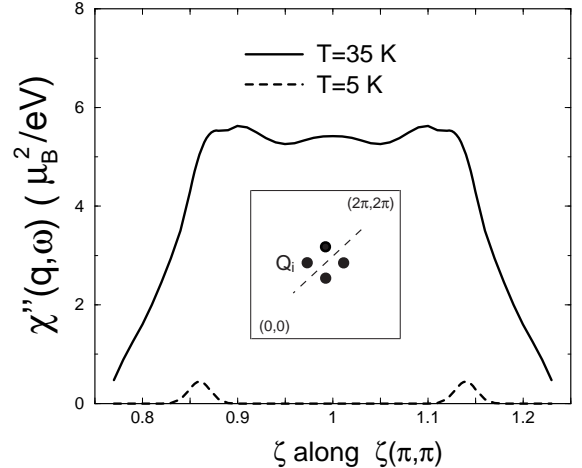}
\end{center}
\caption{$\chi''({\bf q}, \omega)$ in the normal (solid line) and
superconducting  state (dashed line) along the momentum path shown
in the inset.}
\label{chi_qp}
\end{figure}

In Fig.~\ref{chi_qp} we present $\chi_{SC}''$ along the momentum
path shown in the inset for $\omega=8 meV$ in the normal and
superconducting state.
We find that $\chi_{SC}''$ is considerably reduced from its normal
state value along the whole momentum path. This result seems
surprising for ${\bf q}_n$ since here the spin-gap vanishes, and
one might expect comparable intensities in the normal and
superconducting state. We find, however, that due to the
superconducting coherence factors which appear in Eq.(\ref{PiSC}),
Im$\, \Pi_{SC}$ at ${\bf q}_n$ is strongly reduced from its normal
state value, resulting in a decreased intensity in the SC state.

Finally, we note that the form of the spin-gap presented in
Figs.~\ref{om_c1_c2} and \ref{om_c1_comp} as well as the frequency
dependence of $\chi''_{SC}$ for $\omega<\omega_c^{(2)}$ is solely
determined by the cold quasiparticle excitations in the vicinity of
the superconducting nodes. Only for $\omega>\omega_c^{(2)}$ do
hot quasiparticle excitations in the vicinity of $(0,\pi)$ contribute to the
frequency dependence of $\chi''_{SC}$.

\subsection{Magnetic Coherence in YBa$_2$Cu$_3$O$_{6+x}$}
\label{YBCO}

We showed in the previous section that the {\it magnetic coherence
effect} arises from an interplay of an incommensurate magnetic
response and a superconducting gap with $d-$wave symmetry. Since
Dai {\it et al.}~\cite{Dai98,Mook98} recently reported an
incommensurate magnetic structure in the odd spin-channel of
various underdoped YBa$_2$Cu$_3$O$_{6+x}$ compounds, we expect that a similar
effect can be measured for these materials. In the following we consider for
definiteness  YBa$_2$Cu$_3$O$_{6.6}$, the doping level for which the
incommensuration
is the best studied among the YBa$_2$Cu$_3$O$_{6+x}$ compounds.

Due to the bi-layer structure in YBa$_2$Cu$_3$O$_{6+x}$, the
electronic excitations possess a bonding and anti-bonding band,
whose dispersion is described by
\begin{eqnarray}
\epsilon^{a,b}_{\bf k} &=& -2t \Big( \cos(k_x) + \cos(k_y) \Big)
\nonumber \\ & & \quad -4t^\prime \cos(k_x) \cos(k_y) \pm t_\perp
-\mu \ ,
\label{bidisp}
\end{eqnarray}
where $t_\perp$ is the hopping element between nearest neighbors
in adjacent planes. Both bands were recently observed by Schabel
{\it et al.}~\cite{Scha98} in ARPES experiments on optimally doped
YBCO.
\begin{figure} [t]
\begin{center}
\leavevmode
\epsfxsize=7.5cm
\epsffile{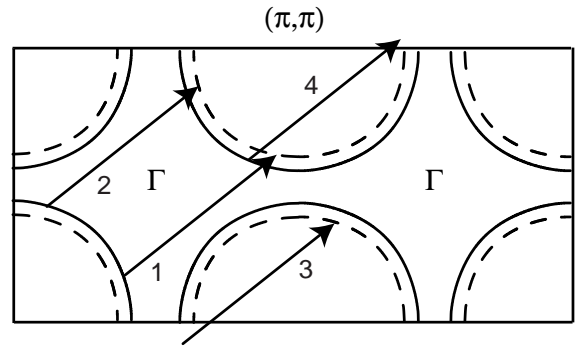}
\end{center}
\caption{The FS of YBa$_2$Cu$_3$O$_{6.6}$. The four arrows
indicate the low-frequency  particle-hole excitations in the odd
channel.}
\label{FS_YBCO}
\end{figure}
From a fit of Eq.(\ref{bidisp}) to the experimentally measured FS
we find $t^\prime/t=-0.5$, $t_\perp/t=0.3$, and $\mu/t =-1.2$; the
corresponding FS is shown in Fig.~\ref{FS_YBCO}. The spin-damping
in the odd channel of $\chi''$ arises from particle-hole
excitations between these two bands which for an incommensurate
wave-vector ${\bf Q}_i$ with $\delta=0.21$ \cite{Mook98} are shown
as arrows in Fig.~\ref{FS_YBCO}.

\begin{figure} [t]
\begin{center}
\leavevmode
\epsfxsize=7.5cm
\epsffile{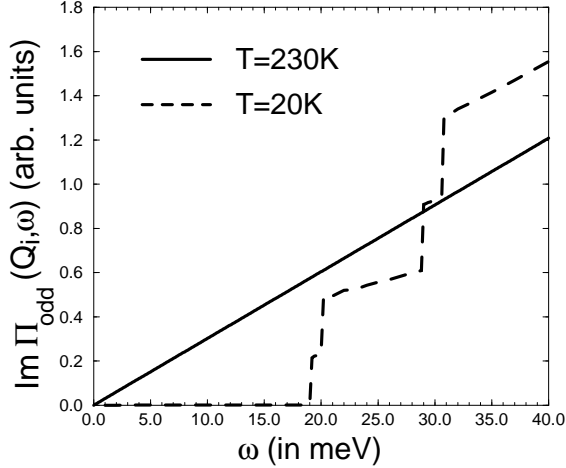}
\end{center}
\caption{Im $\Pi$ at ${\bf Q}_i$ as a function of frequency in the
normal  (solid line) and superconducting state (dashed line).}
\label{sd_YBCO_inc}
\end{figure}

In the normal state these excitations give again rise  to Im$\,
\Pi_{N} \sim \omega$, as is shown in Fig.~\ref{sd_YBCO_inc} (solid
line).
The calculation of Im$\, \Pi$ in the superconducting state of
underdoped YBCO, however, is more complex. We showed in
Sec.~\ref{LSCO} that the magnitude of the spin-gap in the
superconducting state depends on the overall scale, $\Delta_0$, of
the superconducting gap. ARPES experiments in underdoped Bi-2212
compounds \cite{arpes2} have observed that while the gap in the
vicinity of the nodes can still be described by the form given in
Eq.(\ref{dwave}), it acquires an additional momentum dependence in
the vicinity of $(0,\pi)$ and symmetry related points.  Since the
deviation from a pure d-wave symmetry is ascribed to the
appearance of a pseudo-gap in the underdoped cuprates, a similar
effect is to be expected in underdoped YBCO. However, since
excitations 1 and 2 which are predominantly responsible for the magnetic
coherence effect, lie in the
vicinity of the nodes, we can assume that Eq.(\ref{dwave}) still
holds. From the observation by Mook {\it et
al.}~\cite{Mook98,Mookpc} that $\chi_{SC}''$ at ${\bf Q}_i$
disappears in the superconducting state below the spin-gap
frequency $\omega \approx 18$ meV, we infer $\Delta_0 \approx 17$
meV which is consistent with the BCS estimate appropriate to quasiparticles,
$\Delta_0 = 3.5 k_B
T_c$. However, the thresholds defined by excitations 3 and 4
might substantially deviate from the predictions based on the form of
the superconducting gap, Eq.(\ref{dwave}), and in particular are
expected to be larger.

In Fig.~\ref{sd_YBCO_inc} we present Im$\, \Pi_{SC}$ at ${\bf
Q}_i$ as a function of frequency (dashed line). Since the
particle-hole excitations relevant for the odd spin-channel are
inter-band transitions, the four threshold frequencies are
non-degenerate in the superconducting state. This result is
contrary to that obtained for La$_{2-x}$Sr$_x$CuO$_4$ where the
two lower (upper) threshold energies are degenerate. Note,
however, that the energy difference between excitations (1) and
(2), as well as between excitations (3) and (4) is small. Since the momentum
dependence of the superconducting gap is not
known for underdoped YBCO, we
assumed for the calculation of the thresholds that
Eq.(\ref{dwave}) holds for all momenta. As noted earlier, any deviation from
the theoretically predicted values for
the two upper thresholds is a fingerprint of an existing
pseudo-gap. In particular, we predict that the pseudo-gap shifts
the values of the two upper thresholds to higher energies.

\begin{figure} [t]
\begin{center}
\leavevmode
\epsfxsize=7.5cm
\epsffile{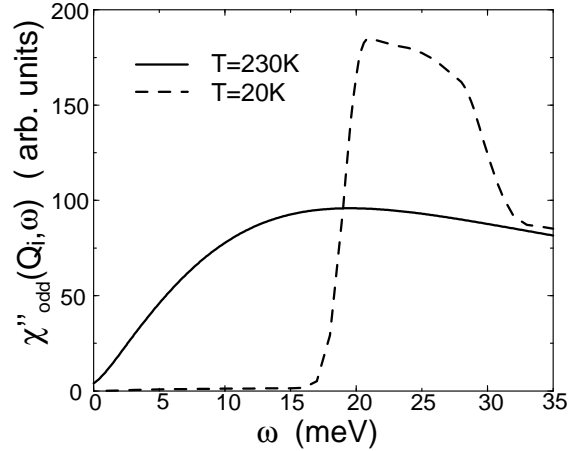}
\end{center}
\caption{$\chi''$ at ${\bf Q}_i$ as a function of frequency in the
normal state  (solid line) and the superconducting state (dashed
line)}
\label{chi_YBCO_inc}
\end{figure}

In  Fig.~\ref{chi_YBCO_inc} we present our theoretical results for
the frequency dependence of $\chi_{SC}''$ at ${\bf Q}_i$
\cite{comm1}.
We again convoluted our theoretical results with a Gaussian
distribution of width $\sigma \approx 2$ meV. In comparing the
strength of the superconducting response to that in the normal
state we chose the normals state response at $T=230$ K to avoid
complications introduced by the strong pseudo-gap behavior.
Similar to our results in Fig.~\ref{sd_om}b
for La$_{2-x}$Sr$_x$CuO$_4$ we find that {\bf (a)} $\chi''_{SC}$
vanishes for frequencies below the spin-gap $\sim 17$ meV, and
{\bf (b)} $\chi''_{SC}$ is increased from its normal state value
at $T=230$ K above $\omega \approx 18$ meV.

\begin{figure} [t]
\begin{center}
\leavevmode
\epsfxsize=7.5cm
\epsffile{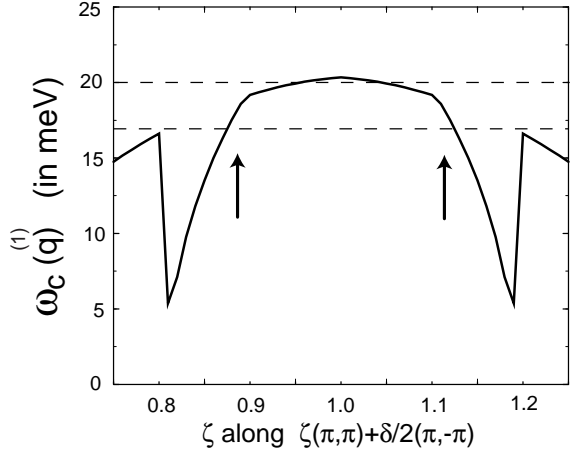}
\end{center}
\caption{Momentum dependence of the spin-gap, i.e., the lowest
frequency threshold, along the path shown in the inset of
Fig.~\protect\ref{sd_om}a. The arrows indicate the positions of
the incommensurate peaks }
\label{spingapYBCO}
\end{figure}

Before we discuss the momentum dependence of $\chi''$ in the
superconducting state, we first consider that of the spin-gap
which we present in Fig.~\ref{spingapYBCO} for the momentum path
shown in the inset of Fig.~\ref{sd_om}a.
It follows from a comparison of Figs.~\ref{om_c1_comp} and
\ref{spingapYBCO} that the bi-layer structure of the FS in YBCO
has a significant impact on the momentum dependence of the
spin-gap. In particular, the spin-gap varies much more strongly
with momentum in YBCO than in LSCO (cf.~Fig.~\ref{om_c1_comp}) from a minimum
$\Delta_{sg}^{min} \approx 5$ meV to a maximum $\Delta_{sg}^{max}
\approx 21$ meV at the midpoint between two incommensurate peaks.

\begin{figure} [t]
\begin{center}
\leavevmode
\epsfxsize=7.5cm
\epsffile{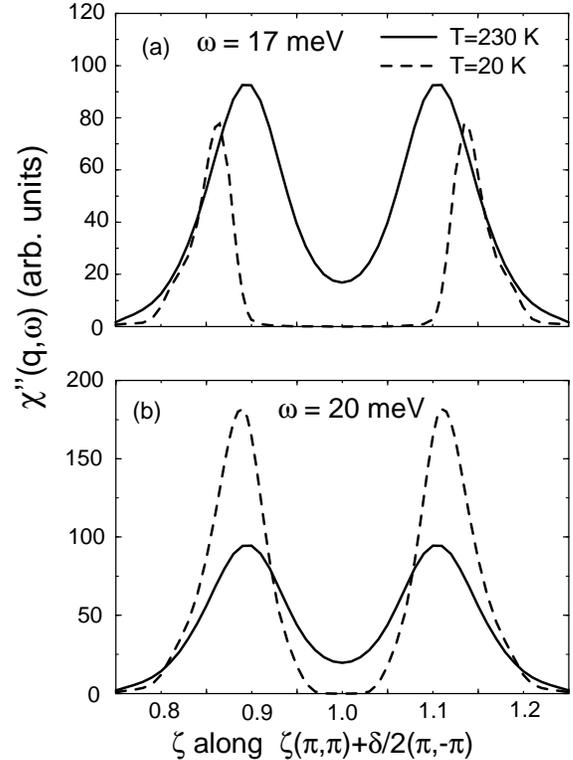}
\end{center}
\caption{  $\chi''({\bf q}, \omega)$ in the normal (solid line)
and superconducting state (dashed line) for {\it (a)} $\omega=17$
meV  and {\it (b)} $\omega=20.0$ meV, along the momentum space
path shown in the inset of Fig.~\protect\ref{sd_om}a.}
\label{chi_YBCO_q}
\end{figure}

In Fig.~\ref{chi_YBCO_q} we present the momentum dependence of
$\chi''$  in the normal and superconducting state for the two
frequencies, $\omega=17$ and 20 meV, denoted by
dashed lines in Fig.~\ref{spingapYBCO}.
As was the case for LSCO, \cite{Morr00a}, $\chi_{SC}''$ for YBCO exhibits a
strong frequency dependence along the momentum path considered
here. For $\omega=17$ meV (Fig.~\ref{chi_YBCO_q}a), the peak
intensity in the superconducting state is anisotropically reduced,
with a stronger suppression of $\chi_{SC}''$ towards the center of
the scan. The anisotropic suppression of $\chi_{SC}''$ is, by
analogy to the case for LSCO, a direct consequence of the momentum
dependence of the spin-gap. Since the spin-gap increases when
moving from ${\bf Q}_i$ toward the center of the scan,
$\chi_{SC}''$ is rapidly cut off by the spin-gap for frequencies
below the maximum spin-gap, $\Delta_{sg}^{max}$. On the other
hand, the spin-gap decreases in the opposite momentum direction,
and the corresponding value for $\chi_{SC}''$ is scarcely reduced. We predict
that this peak
anisotropy should be experimentally observable for all frequencies between
$\Delta_{sg}^{min}$ and $\Delta_{sg}^{max}$. For $\omega=20$ meV
(Fig.~\ref{chi_YBCO_q}b), the anisotropy is reduced and the peak
intensity increases in the superconducting state, as expected from
Fig.~\ref{chi_YBCO_inc}. Since the anisotropy of $\chi_{SC}''$
around ${\bf Q}_i$ is reduced with increasing frequency, the peak
maximum seems to slightly shift towards the center of the scan.

What is the connection between the low energy calculations presented here and
the resonance peak at ${\bf Q}=(\pi,\pi)$ observed at higher
energies ? It is straightforward to extend the low frequency local MMP
expression to higher energies by adding a
spin-wave term in the dynamic spin susceptibility to obtain
\begin{equation}
\chi_0^{-1}={ \xi_0^{-2} + ({\bf q} - {\bf Q})^2 - (\omega/c_{sw})^2 \over
\alpha } \ ,
\label{respeak}
\end{equation}
where $c_{sw}$ is the spin-wave velocity. If then, as
seems plausible, the physical effects responsible for magnetic domain formation
and spin fluctuations are frequency and temperature dependent, then at
sufficiently high frequency, one would expect the domains to disappear. As we
showed in Ref.~\cite{Morr98a}, the form of $\chi_0$ in Eq.(\ref{respeak}) gives
rise to the appearance of a resonance peak in the superconducting state.

\section{Conclusions}
\label{concl}

We have shown that the frequency and momentum dependent changes of $\chi''$ in
the superconducting state of LSCO are a direct consequence of {\bf (a)} an
incommensurate magnetic structure, and {\bf (b)} changes in the
quasi-particle spectrum due to the opening of a superconducting
gap with d-wave symmetry. Our theoretical results for the
frequency dependence of $\chi''$ at ${\bf Q}_i$ are in
quantitative agreement with the available experimental data. We
further find that the momentum dependence of the spin-gap strongly
constrains the form of the Fermi surface, and require a Fermi
surface in La$_{2-x}$Sr$_x$CuO$_4$ which is closed around
$(\pi,\pi)$. It is this sensitivity that is the key to
obtaining information on the electronic excitation spectrum from
INS experiments. We make several predictions for the frequency
dependence of $\chi''(\omega)$ away from ${\bf Q}_i$ that await
further experimental testing. We identify the momentum, ${\bf
q}_n$, in the magnetic BZ that connects the nodes of the
superconducting gap, and for which consequently the spin-gap
vanishes. We discuss the frequency dependence of $\chi_{SC}''$ in
the vicinity of ${\bf q}_n$ and show that even though the spin-gap
vanishes at ${\bf q}_n$, $\chi_{SC}''$ is considerably reduced
from its normal state value. We make
a number of
predictions for the frequency and momentum dependence of
$\chi''$ as well as the form of the spin-gap in
YBa$_2$Cu$_3$O$_{6+x}$, for which an incommensurate magnetic response in the
superconducting state has been seen. The observation of this predicted magnetic
coherence effect in YBa$_2$Cu$_3$O$_{6+x}$ would be an important
step in establishing the universal character of the
magnetic response in the cuprate superconductors.

Finally, to the extent that the predictions made in this
communication are confirmed by experiments, our model would
demonstrate that INS experiments in the superconducting state are {\it not}
confined to
providing information on the magnetic excitation spectrum, but
can
probe as well the {\it electronic spectrum}.
Our model thus makes possible the direct comparison of the results of
INS and ARPES experiments in the superconducting state.

We would like to thank G. Aeppli, R. Birgeneau, A.V. Chubukov, P.
Dai, B. Keimer, B. Lake, T. Mason, A. Millis, H. Mook, and  J.
Schmalian for a large number of stimulating discussions. This work
has been supported by DOE at Los Alamos.


\begin{thebibliography}{99}
\bibitem{comm1} The magnetic correlation length, $\xi$, and the
spin-fermion coupling, $g$, which determine the absolute scale of
$\chi''$ in Fig.~\protect\ref{chi_YBCO_inc} were determined from
our analysis of NMR experiments on YBa$_2$Cu$_3$O$_{6.63}$
\protect\cite{Morr00b}.
\end{thebibliography}
\end{document}